# Exceptional Points in an Specialty Microcavity: Interplay between State-Conversion and Cavity Control Parameters

Arnab Laha, [1] Abhijit Biswas, [2] and Somnath Ghosh [1, *]

[1] *Department of Physics, Indian Institute of Technology Jodhpur, Rajasthan-342037, India*
[2] *Institute of Radio Physics and Electronics, University of Calcutta, Kolkata-700009, India*
*Corresponding author: somiit@rediffmail.com

**Abstract:** Exploiting scattering-matrix in a gain-loss assisted optical-microcavity, interplay between asymmetric-state-conversion and cavity-control parameters around exceptional points is analyzed; where occupying a least area by coupled states during switching, maximum conversion-efficiency with minimal asymmetry is achieved.

## 1. Introduction

Over the last decade, the versatile analogies between non-Hermitian quantum systems and counterpart open optical geometries with suitable amount of simultaneous gain and loss have been attracted surging research attention towards appearance of branch point extreme singularities, namely exceptional points (EPs). Essentially EPs are appeared as topological defect in parameter space where both eigenvalues and corresponding eigenstates of the underlaying system simultaneously coalesce. The presence of an EP leads to crucial modifications in the dynamics of the interacting eigenstates under the influence of coupling parameters. With accurate encirclement around a second order EP in parametric space with adiabatic variation of coupling parameters along a closed contour, one of the associated eigenvalues can permute with its coupled counterpart (i.e. exchanging their positions) in eigenvalue plane [1-5]. This flip-of-states phenomenon plays a key role in the context of asymmetric state-conversion. Such topological characteristics of an EP can be tuned using system parameters viz. geometrical parameters, amount of gain-loss, etc. Appearance of EP with associated unique behaviors have been explored in various open optical systems like optical microcavities [1,2], partially pumped optical waveguides [3,4], laser systems [5], etc.

In this paper, maintaining PT-symmetric constraints, a fabrication feasible balanced gain-loss assisted optical microcavity with non-uniform background refractive index is reported; where widths of background layers are tunable along transverse direction. The interaction phenomenon between the resonances is entirely controlled by two variable parameters of the microcavity viz. amount of gain-loss in terms of gain-coefficient and width of the background layers; where an EP should be encountered without breaking PT-symmetry. To explore an EP, such a typical combination of topological parameters is exploited for the first time to the best-of-our-knowledge. Once an EP is encountered, we analyze the dynamics of the coupled states around EP in the context of asymmetric state conversion and specially focus on how to achieve maximum efficiency in conversion with minimal asymmetry.

## 2. Modelling the microcavity and dynamics of coupled states:

A $1D$ two port open Fabry-Perot type optical cavity has been modeled where an air layer is present between two same high indexed substrates. A spatial distribution of balanced gain-loss profile (with refractive indices as $n_G$ and $n_L$ respectively) is imposed (with strict restriction of PT-symmetric constraints) in terms of gain coefficient $\gamma$ at the high indexed regions as depicted in Fig.1(a). Now to encounter EP without breaking PT-symmetry we have introduced a new tunable parameter $w(x)$ (in $nm$ scale) which essentially fixes the region of occupancy of the intermediate air layer across the cavity cross-section (schematically shown by dotted magenta line). During operations, cavity is accompanied by the poles of the corresponding scattering matrix (defined in Fig.1(a)) which are analogous to the complex eigenvalues of the associated non-Hermitian Hamiltonian. Now obeying current conservation and causality condition, the appeared solutions of the equation $1/\max[eig\,S(\omega)] = 0$ in the lower half of the complex frequency ($k$) plane gives the definite poles of the defined $S$-matrix [1,2,6]. Now, we deliberately choose a specific pair of $S$-matrix poles in a specified frequency range; where with introduction of balanced gain-loss in terms of $\gamma$ (ranges from 0 to 0.03), they are mutually coupled. In Fig. 1(b) we study the avoided resonance crossing (ARC) between them; where abrupt behavioral change in ARCs for $w = 11.6\,nm$ (upper panel) and $w = 11.7\,nm$ (lower panel) confirms the presence of an EP [1-3] in $(\gamma, w)$-plane at ~ (0.0193, 11.65).

Now, to study the intriguing physical effects near/ around the identified EP towards asymmetric state conversion, we consider the effect of encircling around it with a closed circular contour having radius $a$ as shown in Fig. 1(c). Here, three contours are shown with three different $a$-values for clear visibility; while we perform nine encirclements individually around the identified EP with nine choices of $a$, in the range from 0.02 to 0.1 (in a.u.), to study the specific relationship of dynamics of the coupled states around EP with contour parameters. Let, consider $a = 0.04$ (green contour in Fig. 1(c)); where interestingly with one round encirclement around the EP in parameter plane, the corresponding coupled poles exchange their positions in complex $k$-plane along green dotted trajectory as shown in Fig.1(d) i.e. after two successive rounds around EP results in regain of their initial positions forming a complete loop in $k$- plane after second permutation; exhibiting EP as a second order branch point. Now for other $a$-values, the coupled poles follow the similar trajectories as depicted in Fig. 1(d). We calculate the conversion efficiencies using the overlap integrals between two eigenstates [3] associated with the pair of coupled poles; while $1^{st}$ pole switches to second one ($C_{12}$) and vice-versa ($C_{21}$) for each $a$-values. As can be seen in Fig. 1(e), both $C_{12}$ and $C_{21}$ decrease exponentially; where associated asymmetry ($A_C$) grow up exponentially with increase in $a$ (shown in Fig. 1(f)). We also study the dependency of total area ($\Delta$) traversed by pair of coupled poles in $k$-plane during state-flipping w.r.t. the variation in $a$; where similar exponential dependence of $\Delta$, such as $A_C$ with $a$ is displayed in Fig. 1(g). The linear dependence of $A_C$ with $\Delta$, as shown in Fig. 1(h) exhibit the pertinent behavior of them w.r.t. $a$.

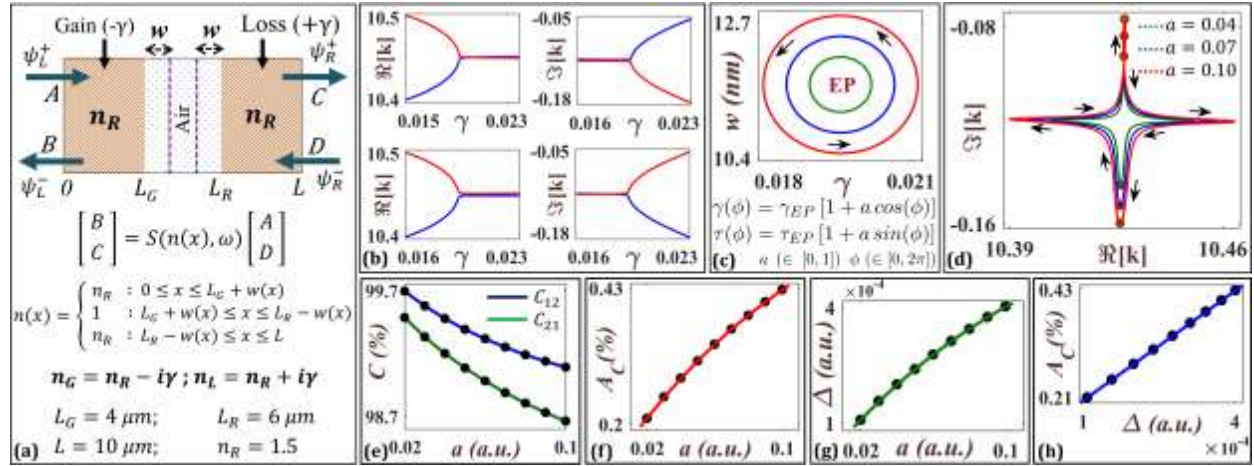

**Fig. 1:** (a) Schematic of the microcavity with all operating parameters and associated scattering matrix where $A$ and $D$ are the complex amplitude corresponding to the incident waves; whereas $B$ and $C$ are complex amplitudes corresponding to the scattered waves. (b) Dynamics of the complex poles w.r.t. $\gamma$; showing anticrossing in $\Re(k)$ and crossing in $\Im(k)$ for $w = 11.6\,nm$ (upper panel), while vice-versa for $w = 11.7\,nm$ (lower panel) respectively. (c) Chosen contours in $(\gamma, w)$-plane with center at respective EP for different values of $a$ and (d) associated dynamics of the interacting poles exhibiting flip-of-sates in $k$-plane. Each contour with different colors in (c) corresponds to each trajectory with respective color in (d). Arrows indicate the direction of evolution. (e) Exponential dependence of $C_{12}$ and $C_{21}$ with $a$ and (f) simultaneous variation of $A_C$. (g) Exponential variation of $\Delta$ w.r.t. $a$ and (h) linear dependence of $A_C$ with $\Delta$.

## 4. Summary

In summary, we report the described specially configured PT-symmetric gain-loss assisted optical microcavity with associated $S$-matrix; where an EP is embedded in cavity parameter plane without breaking PT-symmetry, via interaction between a specified pair of $S$-matrix poles which is entirely controlled by two topological parameters gain-coefficient ($\gamma$) and width of background layers ($w$). A robust mechanism of asymmetric state conversion is reported with a moderately slow variation in $\gamma$ and $w$ along a closed contour around it. We explicitly highlight the specific relationship between state conversion with contour parameters; which revels that to get the maximum conversion with minimal asymmetry, one should trace the location as much as close to the EP in parameter plane for which associated coupled poles traverse the minimum area in complex plane during state-flipping. Such rich physical aspects near EP open up a vast platform to fabricate on-chip state-of-the-art integrated photonic devices.

AL and SG acknowledges support from DST, India [IFA-12; PH-23]. SB acknowledges support from MHRD.